\documentclass[aps,amsmath,amssymb,prb,twocolumn,showpacs,floatfix]{revtex4}
\usepackage{bm}
\usepackage{epsfig}

\newcommand{\cH}{{\mathcal H}}
\newcommand{\br}{{\bf r}}
\newcommand{\be}{{\bf e}}

\newcommand{\ba}{{\bf a}}
\newcommand{\bA}{{\bf A}}
\newcommand{\bF}{{\bf F}}
\newcommand{\bK}{{\bf K}}
\newcommand{\bG}{{\bf G}}
\newcommand{\bX}{{\bf X}}
\newcommand{\bS}{{\bf S}}
\newcommand{\bC}{{\bf C}}
\newcommand{\bq}{{\bf q}}
\newcommand{\bk}{{\bf k}}
\newcommand{\bkaf}{{\bf k}_{\rm AF}}
\newcommand{\bQ}{{\bf Q}}

\newcommand{\qch}{{\bf Q}^{\textrm{ch}}}
\newcommand{\bM}{{\bf M}}

\renewcommand{\S}{{\bf S}}
\newcommand{\nn}{{\nonumber}}
\newcommand{\1}{{(1)}}
\newcommand{\2}{{(2)}}
\newcommand{\Sx}{{\tilde S^x}}
\newcommand{\Sy}{{\tilde S^y}}
\newcommand{\Sz}{{\tilde S^z}}
\newcommand{\Sp}{{\tilde S^+}}
\newcommand{\Sm}{{\tilde S^-}}
\newcommand{\bd}{b^\dagger}

\newcommand{\llangle}{\langle\kern-.25em\langle}
\newcommand{\rrangle}{\rangle\kern-.25em\rangle}
\newcommand{\w}{\omega}

\newcommand{\cS}{{\mathcal S}}
\newcommand{\fin}{\textrm{F}}
\newcommand{\dk}{\delta k}

\newcommand{\coma}{cond-mat/}

\begin{document}

\title{The spin excitation spectrum in striped bilayer compounds}

\author{Frank Kr\"uger}
\author{Stefan Scheidl}

\affiliation{Institut f\"ur Theoretische Physik, Universit\"at zu
  K\"oln, Z\"ulpicher Str. 77, D-50937 K\"oln, Germany}

\begin{abstract}
  The spin dynamics of bilayer cuprate compounds are studied in a
  basic model.  The magnetic spectral properties are calculated in
  linear spin-wave theory for several stripe configurations which
  differ by the relative location of the stripes in the layers.  We
  focus on the bilayer splitting of the magnon bands near the
  incommensurate low energy peaks as well as near the $\pi$ resonance,
  distinguishing between the odd and even channel.  We find that a
  x-shaped dispersion near the $\pi$ resonance is generic for stripes.
  By comparison of our results to neutron scattering data for
  $\mathrm{YBa_2Cu_3O_{6+x}}$ we conclude that the stripe model is
  consistent with characteristic features of bilayer high-$T_c$
  compounds.
\end{abstract}
\date{\today}

\pacs{75.10.Jm, 74.72.-h, 75.30.Ds, 76.50.+g} \maketitle

\section{Introduction}
\label{sec.intro}

Subsequent to predictions of stripe
formation,\cite{Zaanen+89,Schulz89,Machida89} characteristic
signatures of spin- and charge order have been found in a variety
of high-$T_c$ cuprate superconductors, including
$\mathrm{La_{2-x}Sr_xCuO_4}$ (LSCO) and
$\mathrm{YBa_2Cu_3O_{6+x}}$ (YBCO). Neutron scattering
experiments\cite{Cheong+91,Mook+98} have provided evidence for
spin order at low energies through a pattern of incommensurate
peaks around the antiferromagnetic wave vector. Although more
difficult to detect, charge order has been observed in LSCO
co-doped with $\mathrm{Nd}$\cite{Tranquada+95} as well as in YBCO
without codoping.\cite{Mook+02}

Since LSCO and YBCO are paradigmatic for monolayer and bilayer
compounds, stripe-like ``low''-energy response is characteristic for
both classes of materials.  On the other hand, at ``high'' energies
spin fluctuations appeared to be qualitatively different since a
commensurate $\pi$-resonance had been observed only in bilayer
compounds, notably in
$\mathrm{YBa_2Cu_3O_{6+x}}$\cite{Rossat-Mignod+91,Fong+95} and
$\mathrm{Bi_2Sr_2CaCu_2O_{8+x}}$,\cite{Keimer+99} whereas it seemed to
be absent in monolayer compounds.  This apparent distinction between
mono- and bilayer compounds lost its justification only recently, when
the $\pi$-resonance was discovered in
$\mathrm{Tl_2Ba_2CuO_{6+x}}$\cite{He+02} as the first monolayer
compound.  The fact that the $\pi$-resonant mode has not been detected
in LSCO so far can possibly be ascribed to a larger effective strength
of disorder, since the $\mathrm{Sr}$-dopants are randomly distributed
whereas in the oxygen doped compounds the access oxygen orders in
chains. Thus, one may believe that, in principle, mono- and bilayer
compounds have qualitatively similar features also at higher energies.
This universality of low- and high-energy features calls for an even
more unifying framework.\cite{Batista+01}

In a recent article\cite{Kru+03} we have analyzed an elementary
monolayer model assuming that charges form a perfectly ordered
site-centered stripe array which imposes a static spatial modulation
of spin-exchange couplings.  The resulting spin dynamics was studied
using linear spin-wave theory.  As a result, we found that the
incommensurability and the $\pi$-resonance appear as complementary
features of the band structure at different energy scales.
Furthermore, the doping dependence of the resonance frequency was
found in good agreement with experimental observations.

In this work we extend this model to bilayer systems in order to
predict the corresponding features of the magnon band structure and
the magnetic structure factor.  Within each layer, holes are assumed
to form unidirectional site-centered stripes. We consider several
possibilities (parallel and perpendicular relative orientations) of
the charge order in the antiferromagnetically coupled neighboring
layers. The band structure and the $T=0$ inelastic structure factor
for even and odd excitations are calculated in linear spin-wave
theory. Particular attention is paid to the band-splitting in the
vicinity of the antiferromagnetic wave vector and to the influence of
the interlayer coupling on the $\pi$-resonance energy.

The outline of this paper is as follows. In Sec.~\ref{sec.model} the
spin-only model for a bilayer system is introduced and motivated.
Classical ground states and the resulting phase diagrams for competing
types of magnetic order are obtained.  They are needed as starting
point for the linear spin-wave theory.  A customized formulation
thereof is outlined in Sec.~\ref{sec.sw}. The results, namely the
spin-wave band structure, the zero-temperature structure factor for
even and odd excitations, and the dependence of the band splitting at
the antiferromagnetic wave-vector on the strength of the interlayer
coupling are presented in Sec.~\ref{sec.results} and compared to
experiments in Sec.~\ref{sec.discussion}.

\section{Model}
\label{sec.model}

Stripes are a combined charge- and spin-density wave.  If the charge
period $pa$ is a multiple of the Cu spacing $a$ with integer $p$,
lock-in effects tend to suppress phason-like fluctuations of the
density modulation.  In a reductionist real-space picture, one may
think of the holes forming parallel site-centered rivers of width $a$,
which act as antiphase domain boundaries for the antiferromagnetic
spin domains in between.\cite{Tranquada+95} This implies that the
period of the spin modulation is twice that of the charge modulation.

To implement that the charge stripes act like antiphase boundaries we
follow our previous work\cite{Kru+03} and choose the simplest possible
implementation of exchange couplings within the layers stabilizing
this magnetic structure: antiferromagnetic exchange couplings $J$
between neighboring spins within the domains and antiferromagnetic
couplings $\lambda J$ between closest spins across a stripe.

\begin{figure}[h]
\epsfig{figure=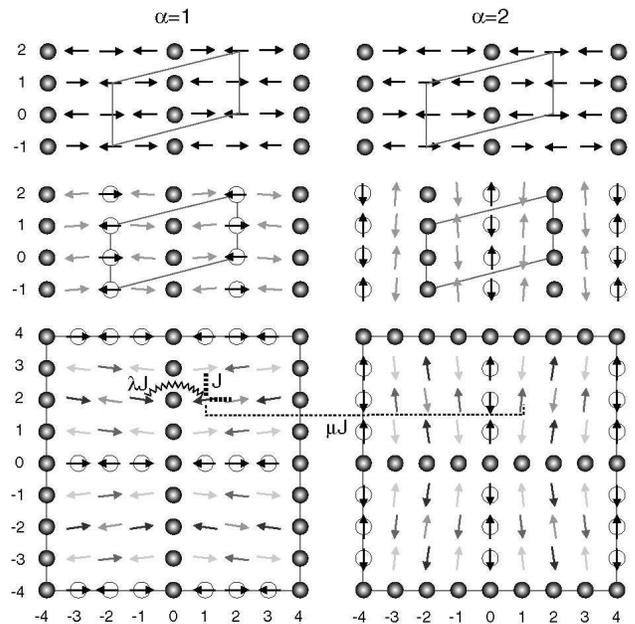,width=0.95\linewidth}
\caption{%
  Classical ground states for bilayer systems with parallel (upper
  row), shifted parallel (middle row), and perpendicular hole stripes
  (lower row) for a stripe spacing $p=4$. The exchange couplings of
  the simple model are illustrated in the lower row: AF couplings
  $J>0$ for nearest neighbors within the domains (bold dashed),
  $\lambda J$ between nearest neighbors across a hole stripe (zig-zag)
  and couplings $\mu J$ between spins one above the other (dashed).
  Frustration of exchange coupling may lead to a canting of spins
  (calculated for $\mu=0.09$ and $\lambda=0.07$ in the middle and
  bottom row, respectively).  Possible magnetic unit cells are
  outlined by gray lines, identical gray levels of spins correspond to
  identical canting angles.}
  \label{gs}
\end{figure}

In our previous work\cite{Kru+03} we have studied this model for a
single layer allowing for diagonal and vertical stripe orientations.
Here we focus on vertical stripes as observed in the superconducting
cuprates and restrict our analysis to the representative case $p=4$.
This corresponds to a doping of one hole per 8 Cu sites since the
rivers have a line charge of only half a hole per lattice constant.
In addition to the in-plane couplings we consider an antiferromagnetic
exchange $\mu J$ between two layers (cf.  Fig.~\ref{gs}). The
Hamiltonian of this bilayer model is given by
\begin{subequations}
\begin{eqnarray}
  H & = & \sum_{\alpha=1,2}H_\alpha+H_{1,2},
  \\
  H_\alpha & = & \frac{1}{2}\sum^*_{\br,\br'}
  J_\alpha(\br,\br')\S_\alpha(\br)\S_\alpha(\br'),
  \\
  H_{1,2}& = & \mu J \sum^*_\br \S_1(\br)\S_2(\br),
\end{eqnarray}
\end{subequations}
where $\br$ specifies the square-lattice position and $\alpha=1,2$
numbers the layers.  The asterisks indicate that the sums do not
include positions of charge rivers. The in-plane couplings
$J_\alpha(\br,\br')$ defined in the text above are illustrated in
Fig.~\ref{gs}.  They explicitly depend on the layer index if the
charge distribution is different in both layers.

For simplicity we neglect spin anisotropy, the weak 3D coupling
between bilayers, and more complicated exchange processes such as
cyclic exchange or Dzyaloshinskii-Moriya interactions, which all
may be important for quantitative purposes.  Obviously, this
simple spin-only model does not account for electronic correlation
effects, e.g. a spin gap at low energies due to the formation of
Cooper pairs is not incorporated.  Nevertheless we expect that our
model provides a qualitatively adequate description of the spin
fluctuations well above the gap energy.

The actual stripe configuration is determined by several influences.
Besides the magnetic exchange energy one also has to take into account
the Coulomb energy, and in principle also a further reduction of the
fourfold symmetry of CuO$_2$ planes in orthorhombic structures which
may favor a certain alignment of the stripes.  In
$\mathrm{YBa_2Cu_3O_{6+x}}$ the formation of $\mathrm{CuO}$ chains
along the $b$-direction may favor a parallel alignment of stripes.

We find that three different stripe configurations may be realized
physically (see Fig.~\ref{gs}).  The exchange energy favors
\textit{parallel stripes} lying exactly on top of each other.  This
configuration is free of magnetic exchange frustration, each bond can
be fully saturated.  However, this configuration is disfavored by the
Coulomb energy which would favor a configuration where stripes are
\textit{parallel but shifted} with respect to each other by half a
stripe spacing. (In our simple model, where holes are assumed to be
site centered, this configuration is only compatible with even stripe
spacings $p$.)  The gain in Coulomb energy has to be paid by a loss of
exchange energy.  For certain parameters, a third configuration may be
favorable, where the charge stripes of the two layers are
\textit{perpendicular}.

For the later analysis it is instructive to anticipate that for
these configurations the Hamiltonian has discrete symmetries.  We
focus on symmetries involving an exchange of layers.  For parallel
and shifted parallel stripes, this symmetry is just the reflection
$z \to - z$ combined with a translation (coordinates are chosen
such that the planes are parallel to the $xy$ plane).
 For perpendicular stripes, one needs to
add a rotation around the $z$ axis.

\subsection{Energetic estimates}

To estimate the Coulomb energy for the three stripe configurations, we
assume a charge-density modulation $\rho(\br) = \rho_1(\br)\delta(z) +
\rho_2(\br)\delta(z-d)$ with $\rho_\alpha(\br)=\rho_\alpha^{(0)}
\cos(\bk_\alpha\br)$ where the planes separated by $d$ are
perpendicular to the $z$-direction.  For simplicity, only the first
harmonic of the charge modulation is retained.  Parallel stripes are
described by $\bk_1=\bk_2=k \be_x$ and $\rho_1^{(0)} = \rho_2^{(0)} =
\bar{\rho}$, shifted parallel stripes are realized for $\bk_1=\bk_2=k
\be_x$ and $\rho_1^{(0)} = -\rho_2^{(0)} = \bar{\rho}$, and
perpendicular stripes for $\bk_1=k \be_x$, $\bk_2=k \be_y$ and
$\rho_1^{(0)} = \rho_2^{(0)} = \bar{\rho}$. For a stripe spacing $pa$
the charge-modulation wave vectors are given by $k=2\pi/(pa)$, the
amplitude by $\bar{\rho}=e/(2pa^2)$.  Calculating the Coulomb coupling
energy per square lattice site
\begin{equation}
  E_C=\frac{1}{4\pi\epsilon_0}\frac{a^2}{A}
  \int\textrm{d}^3r\int\textrm{d}^3r'
  \frac{\rho_1(\br)\rho_2(\br')}{|\br-\br'|},
\end{equation}
where $A$ denotes the area of the planes, we find, in the limit
$A\to\infty$, a vanishing Coulomb coupling for perpendicular
stripes, an energy cost
\begin{equation}
  E_{C}= \frac{e^2}{32\pi\epsilon_0pa}
  \exp\left(-2\pi\frac{d}{pa}\right)
\end{equation}
for parallel stripes, and an energy gain of the same size for
shifted parallel stripes.  For YBCO with $a\approx 3.85$~\AA,
$d\approx 3.34$~\AA, $J=125$~meV, $S=\frac 12$ and for a stripe
spacing $p=4$ we obtain $\Delta E_{C}\approx 29$~meV.

For antiferromagnetic YBCO the magnetic interlayer superexchange
is reported to be $\mu \approx 0.08$.\cite{Reznik+96} For parallel
stripes, spins are not frustrated and, in a classical picture,
antiparallel in different layers, $\S_1(\br)=-\S_2(\br)$. Thus,
the exchange coupling roughly leads to an energy gain of order
$\mu J S^2 \approx 3$~meV, whereas the energy gain will be smaller
for the other two configurations due to frustration.

Thus, within our rough estimate, the Coulomb energy appears to be
up to one order of magnitude larger than the exchange energy, such
that one might expect the parallel shifted configuration to be the
only physical one. On the other hand the actual Coulomb energy may
be significantly smaller than the result of our estimate since we
have completely neglected screening. For almost undoped YBCO a
relative large value of $\epsilon\approx 15$ for the static
dielectric constant at $T=4$K is reported.\cite{Samara+90}
Therefore the Coulomb energy might be of the same order of
magnitude as the magnetic exchange energy.  Due to the crudeness
of our estimate no stripe configuration can be strictly ruled out.

\subsection{Classical ground states}

Due to frustration effects, the ground-state structure is nontrivial
for shifted parallel and perpendicular stripes. We now determine these
ground states treating spins as classical.  These ground states will
be a necessary prerequisite for the subsequent spin-wave analysis.  We
continue to focus on the representative case $p=4$.

\begin{figure}[h]
\epsfig{figure=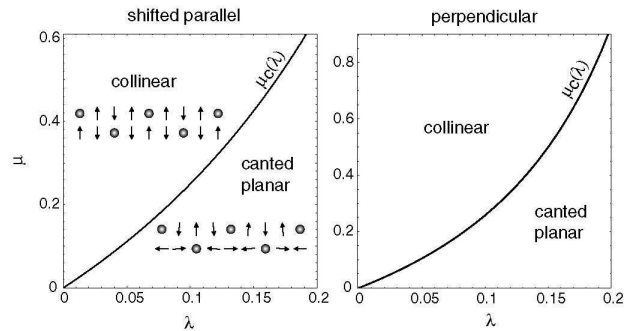,width=0.95\linewidth}
\caption{%s
  Classical ground-state phase diagrams for shifted parallel and
  perpendicular stripes. For $\mu<\mu_c(\lambda)$ the ground states
  show a canted planar spin pattern illustrated in Fig.~\ref{gs}.  For
  $\mu>\mu_c(\lambda)$ the topology of the ground states changes into
  a collinear pattern where spins lying on top of each other are
  strictly antiparallel and nearest neighbors across a stripe are
  parallel. For $\lambda\to\lambda_c$ ($\lambda_c\approx 0.59$ for
  shifted parallel and $\lambda_c\approx 0.35$ for perpendicular
  stripes) $\mu_c$ goes to infinity. Above $\lambda_c$ the ground
  states are always planar.}
\label{mulam}
\end{figure}

Depending on the values of the couplings $\lambda$ and $\mu$ we
find two different types of ground states. For a nearest neighbor
exchange across a stripe in the range $0<\lambda<\lambda_c$
($\lambda_c\approx 0.59$ for shifted parallel and
$\lambda_c\approx 0.35$ for perpendicular stripes) the ground
state has a canted planar topology up to a value $\mu_c(\lambda)$
of the interlayer exchange (cf. Fig.~\ref{mulam}). For
$\mu>\mu_c(\lambda)$ spins lock into a collinear texture.

\begin{figure}[h]
\epsfig{figure=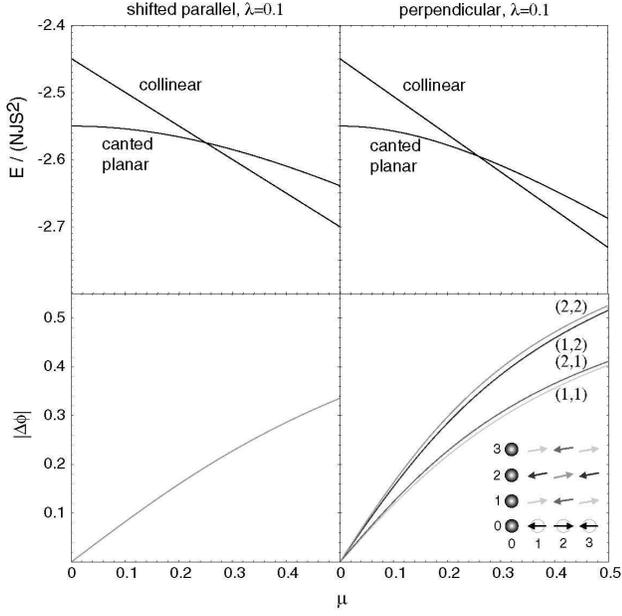,width=0.95\linewidth}
\caption{%
  Upper row: Energy per lattice site in units of $JS^2$ as a function
  of $\mu$ for $\lambda=0.1$. For both stripe configurations the
  energies of the canted planar and the collinear spin pattern are
  plotted.  The curves intersect at $\mu=\mu_c$ where the topology of
  the ground states changes.  Lower row: Relative values of the
  tilting angles of the spins in the planar configuration as a
  function of the interlayer coupling $\mu$ for $\lambda=0.1$.}
 \label{angene}
\end{figure}

To characterize these different phases, we start with the planar
one. As already indicated above, the frustration can lead to a
canting of spins.  The origin of the canting is easily understood.
For $\mu=0$ the layers are decoupled and the sublattice
magnetization in both layers can have an arbitrary relative
orientation.  For small interlayer coupling $\mu$ the spins start
to cant starting from a configuration where spins lying on top of
each other are perpendicular.  Only in this case the interlayer
couplings lead to an energy gain proportional to small canting
angles while the intralayer couplings lead to an energy costs of
second order in the canting angles.  Such canted planar ground
states are illustrated in Fig.~\ref{gs}. In Fig.~\ref{angene} the
corresponding tilting angles are plotted for $\lambda=0.1$ as a
function of $\mu$.  The tilting angles increase monotonously in a
way that spins lying on top of each other become increasingly
antiparallel with increasing $\mu$.

In the other phase, for $\mu>\mu_c(\lambda)$, the interlayer
coupling $\mu$ dominates the coupling $\lambda$ across the stripes
and the topology of the ground state changes into a collinear
configuration where the spins lying on top of each other are
strictly antiparallel and nearest neighbor spins across a stripe
are strictly parallel although they are antiferromagnetically
coupled. This configuration is stable against a canting of the
spins because for small $\lambda$ the energy gain for
$\lambda$-bonds and the energy costs for $\mu$-bonds as well as
the couplings within the domains would be quadratic in the tilting
angles. Since this ground state has lost the antiphase-boundary
character of the charge stripes it resembles a diluted
antiferromagnet.  This would lead to a static magnetic response at
the antiferromagnetic wave vector in disagreement with
experimental observations.  Therefore, these collinear phases
probably are not relevant for the magnetic properties of the
cuprate compounds.

For small values of $\lambda$ the phase boundary is approximately
given by $\mu_c(\lambda)\approx 2\lambda$ for both stripe
configurations (cf. Fig.~\ref{mulam}). In the limit
$\lambda\to\lambda_c$ the critical value $\mu_c$ goes to infinity.
Above $\lambda_c$ the ground states remain canted planar for all
values of the interlayer coupling $\mu$.

Comparing the classical magnetic ground-state energies for the two
frustrated configurations, we find that -- in contrast to the
Coulomb energy -- the exchange coupling favors perpendicular
stripes over shifted parallel stripes.  For this reason we retain
perpendicular stripes in our consideration.

\section{Spin-wave Theory}
\label{sec.sw}

In this analytic part we derive general expressions for the magnon
band structure and the spectral weight at zero temperature in a
framework of linear spin-wave theory (for a review in the context of
cuprates, see e.g. Ref. \onlinecite{Manousakis91}). These expressions
are evaluated numerically later on in Sec.~\ref{sec.results} for
parallel, shifted parallel and perpendicular stripes and fixed stripe
spacing $p=4$.

\subsection{Holstein-Primakoff representation}

The ground-state analysis of the preceding section has made clear
that spin waves now have to be introduced as excitation of a
non-collinear ground state.  However, our numerical calculation of
the classical ground states have shown planar spin textures (here,
a collinear texture is considered as a special subcase of a planar
texture).

In the following we consider a general planar ground state which can
be captured by a vector field $\S_{\alpha}(\br) =
\{\cos\phi_{\alpha}(\br), \sin\phi_{\alpha}(\br),0\}$, where the
tilting angles of the spins obey the translational symmetry
$\phi_\alpha(\br) = \phi_\alpha(\br+\bA)$ for an arbitrary magnetic
lattice vector $\bA = m_1\bA^\1+m_2\bA^\2$.  For the spin textures
displayed in Fig.~\ref{gs}, corresponding magnetic unit cells are
given by $\bA^\1=(4,1)$ and $\bA^\2=(0,2)$ for parallel stripes and
for shifted parallel stripes, and by $\bA^\1=(8,0)$ and $\bA^\2=(0,8)$
for perpendicular stripes.

To study the quantum fluctuation around the classical ground state we
rotate all spins by their planar angles $\phi_{\alpha}(\br)$ according
to
\begin{subequations}
\begin{eqnarray}
  S_\alpha^x(\br) & = & \Sx_\alpha(\br)\cos\phi_\alpha(\br)
  -\Sy_\alpha(\br)\sin\phi_\alpha(\br), \\
  S_\alpha^y(\br) & = & \Sx_\alpha(\br)\sin\phi_\alpha(\br)
  +\Sy_\alpha(\br)\cos\phi_\alpha(\br), \\
  S_\alpha^z(\br) & = & \Sz_\alpha(\br),
\end{eqnarray}
\end{subequations}
such that $\tilde\S(\br)$ has a classical ferromagnetic ground state
$\tilde\S(\br)=S\{1,0,0\}$.  In the transformed spin basis we
introduce Holstein-Primakoff (HP) bosons in the standard way (using
$\tilde S^\pm=\Sy\pm i\Sz$),
\begin{subequations}
\begin{eqnarray}
  \Sp_\alpha(\br) & = & \sqrt{2S-\hat{n}_{\br,\alpha}} \ b_{\br,\alpha},\\
  \Sm_\alpha(\br) & = & \bd_{\br,\alpha}\sqrt{2S-\hat{n}_{\br,\alpha}},\\
  \Sx_\alpha(\br) & = & -\hat{n}_{\br,\alpha}+S,
\end{eqnarray}
\end{subequations}
and obtain the spin-wave Hamiltonian
\begin{eqnarray}
  \cH & = &
  \frac{S}{2}\sum_{\br,\br'}^*\sum_{\alpha,\alpha'}
  \left\{f_{\alpha,\alpha'}(\br,\br')
    \left[\bd_{\br\alpha}b_{\br'\alpha'}
      +b_{\br\alpha}\bd_{\br'\alpha'}\right]\right.\nn
  \\
  & & \phantom{\frac{S}{2}\sum_{\br,\br'}\sum_{\alpha,\alpha'}}
  +\left.g_{\alpha,\alpha'}
    (\br,\br')\left[b_{\br\alpha}b_{\br'\alpha'}
      +\bd_{\br\alpha}\bd_{\br'\alpha'}\right]\right\},\quad
\end{eqnarray}
where the functions $f$ and $g$ are defined by
\begin{subequations}
\begin{eqnarray}
f_{\alpha,\alpha'}(\br,\br') & = & \frac 12
\left[J_\alpha(\br,\br')\delta_{\alpha,\alpha'}+\mu
J\delta_{\br,\br'}\left(1-\delta_{\alpha,\alpha'}\right)\right]\nn\\
& & \times\left[\Delta_{\alpha,\alpha'}(\br,\br')+1\right]\nn\\
&  &
-\delta_{\br,\br'}\delta_{\alpha,\alpha'}\sum_{\br'}J_\alpha(\br,\br')\Delta_{\alpha,\alpha'}(\br,\br')\nn\\
& & -\mu J\delta_{\br,\br'}\delta_{\alpha,\alpha'}\sum_{\alpha'}
(1-\delta_{\alpha,\alpha'})\Delta_{\alpha,\alpha'}(\br,\br')\nn\\
\\
g_{\alpha,\alpha'}(\br,\br') & = & \frac 12
\left[J_\alpha(\br,\br')\delta_{\alpha,\alpha'}+\mu
J\delta_{\br,\br'}\left(1-\delta_{\alpha,\alpha'}\right)\right]\nn\\
& & \times\left[\Delta_{\alpha,\alpha'}(\br,\br')-1\right]\\
\Delta_{\alpha,\alpha'}(\br,\br')& = &
\cos\left[\phi_\alpha(\br)-\phi_{\alpha'}(\br')\right].
\end{eqnarray}
\end{subequations}
To diagonalize the Hamiltonian, we Fourier transform the bosonic
operators via $b_{\alpha}(\br)=\int_\bk \exp(i\bk\br)b_{\alpha}(\bk)$,
where $\int_\bk = (2 \pi)^{-2} \int d^2 k$ and the $\bk$ integrals run
over the Brillouin zone of the square lattice with an area
$(2\pi/a)^2$.  Following our calculations for the monolayer
system\cite{Kru+03} we decompose a square lattice vector $\br$ into a
magnetic lattice vector $\bA$ and a decoration vector $\ba$
($\br=\bA+\ba$). The number of vectors $\ba$ is denoted by $n$ (the
area of the magnetic unit cell).  In momentum space, the reciprocal
magnetic basis $\bQ^{(i)}$, $i=1,2$, spans the corresponding magnetic
Brillouin zone ($\mathcal{BZ}$).  Wave vectors $\bk$ can be uniquely
decomposed into $\bk=\bQ+\bq$ with $\bq\in\mathcal{BZ}$ and
$\bQ=m_1\bQ^\1+m_2\bQ^\2$.  Within the Brillouin zone of the square
lattice there are $n$ vectors $\bQ$ which we denote by $\bQ_\nu$.
Using these decompositions we rewrite the spin-wave Hamiltonian as
\begin{eqnarray}
\label{ham}
  \cH & = & \frac{1}{2}\int_\bq\sum_{\nu,\nu'}\sum_{\alpha,\alpha'}
  F_{\nu\alpha,\nu'\alpha'}(\bq)
  [\bd_{\alpha,\bq+\bQ_\nu}b_{\alpha',\bq+\bQ_{\nu'}}\nn\\
  & & \qquad+b_{\alpha,-\bq-\bQ_\nu}\bd_{\alpha',-\bq-\bQ_{\nu'}}]\nn \\
  &  &+ \frac{1}{2}\int_\bq\sum_{\nu,\nu'}\sum_{\alpha,\alpha'}
  G_{\nu\alpha,\nu'\alpha'}(\bq)
  [\bd_{\alpha,\bq+\bQ_\nu}\bd_{\alpha',-\bq-\bQ_{\nu'}}\nn\\
  & & \qquad+b_{\alpha,-\bq-\bQ_\nu}b_{\alpha',\bq+\bQ_{\nu'}}] ,
\end{eqnarray}
where
\begin{eqnarray}
  F_{\nu\alpha,\nu'\alpha'}(\bq) & = & \frac{S}{n}\sum_{\bA}
  \sum_{\ba,\ba'}f_{\alpha,\alpha'}(\ba+\bA,\ba')
  \nn\\
  & & \times \cos\left[\bq\bA+\bq(\ba-\ba')+\bQ_\nu\ba-\bQ_{\nu'}\ba'\right]
\end{eqnarray}
is essentially the Fourier transform of $f$,
\begin{eqnarray}
  \frac{S}{n} f_{\alpha,\alpha'}(\bQ_\nu+\bq,\bQ_{\nu'} + \bq') =
  \delta(\bq+\bq')  F_{\nu\alpha,\nu'\alpha'}(\bq) .
\end{eqnarray}
Analogous expressions relate $G$ to $g$. The Hamiltonian (\ref{ham})
has exactly the same structure as in the monolayer case [compare Eq.~8
in Ref. \onlinecite{Kru+03}] and can be diagonalized by a Bogoliubov
transformation in an analogous way. The final diagonal form is given
by
\begin{equation}
\cH=\sum_{\gamma=1}^{2n}\int_\bq
\w_\gamma(\bq)\left\{\bd_\gamma(\bq)b_\gamma(\bq)+\frac
12\right\},
\end{equation}
where the squared energies $\w_\gamma^2$ are eigenvalues of the
hermitian matrix $\bM^{-1/2}\bK\bM^{-1/2}$. Thereby
$\bM^{-1}=\bF-\bG$ denotes the inverse mass matrix and
$\bK=\bF+\bG$ the coupling matrix.

\subsection{Structure Factor}

We now proceed to calculate the inelastic zero-temperature structure
factor for even and odd excitations
\begin{eqnarray}
  \cS^\textrm{in}_\pm (\bk,\w) &:=& \sum_\fin \sum_{j=x,y,z}
  |\langle \fin | S^j_1(\bk)\pm S^j_2(\bk)|0 \rangle |^2 \nn\\
  & & \qquad\times\delta (\w-\w_\fin) .
\label{struc}
\end{eqnarray}
Here, $|0\rangle$ denotes the ground state (magnon vacuum)
characterized by $b_\gamma(\bq) |0\rangle =0$ and we consider only
single-magnon final states $|\fin\rangle=\bd_\gamma(\bq)
|0\rangle$ with excitation energy $\w_\fin:=E_\fin-E_0$.
$\bk=(k_x,k_y)$ denotes the in-plane wave-vector, odd excitations
correspond to $k_z^-=(2n+1)\pi/d$ [$L^-=(2n+1)c/(2d)$ in
reciprocal lattice units], even ones to $k_z^+=2n\pi/d$
($L^+=nc/d$), where $d$ is the distance of the two layers within
the orthorhombic unit cell.  For YBCO with $d \approx 3.34$~{\AA}
and $c \approx 11.7$~{\AA} the corresponding values for even and
odd modes are $L^- \approx 1.75, 5.25$ and $L^+ \approx 0, 3.5$.

Expressing the spin operators by the final bosonic operators
$b_\gamma(\bq)$ it is straightforward to calculate the structure
factor. Using a pseudo-Dirac notation and denoting the
$2n$-dimensional cartesian basis by $|\nu,\alpha \rrangle$
($\nu=1,\ldots N, \alpha=1,2$) and the orthonormal eigenbasis of
$\bM^{-1/2}\bK\bM^{-1/2}$ by $|\gamma \rrangle$, the structure factor
can be rewritten in a compact form,
\begin{subequations}
\begin{eqnarray}
  \cS^\textrm{in}_\pm (\bq+\bQ_\nu,\w) &=&  S \sum_{\gamma}
  \cS^\pm_\gamma(\bq+\bQ_\nu)  \delta (\w-\w_\gamma(\bq)),\qquad
  \\
  \cS^\pm_\gamma(\bq+\bQ_\nu)&=&\frac 12\sum_{\bX=\bC,\bS,\w^{-1}_\gamma \bK}
  \llangle \nu,\pm| \bX\bM^{-1/2} |\gamma \rrangle\nn\\
  & & \qquad\times
  \frac 1{\w_\gamma} \llangle \gamma |\bM^{-1/2}\bX |\nu,\pm \rrangle,\qquad
\end{eqnarray}
\label{strucf}
\end{subequations}
where we have defined $|\nu,\pm \rrangle=(1/\sqrt{2})[|\nu,1
\rrangle\pm |\nu,2 \rrangle]$ and introduced the matrices $\bS$ and
$\bC$ according to
\begin{subequations}
\begin{eqnarray}
  s_{\nu\alpha,\nu'\alpha'} & = & \frac
  1n\delta_{\alpha\alpha'}\sum_\ba^*\sin\phi_\alpha(\ba)
  e^{i(\bQ_\nu-\bQ_{\nu'})\ba} ,
  \\
  c_{\nu\alpha,\nu'\alpha'} & = & \frac
  1n\delta_{\alpha\alpha'}\sum_\ba^*\cos\phi_\alpha(\ba)
  e^{i(\bQ_\nu-\bQ_{\nu'})\ba} .
\end{eqnarray}
\end{subequations}

\section{Results}
\label{sec.results}

We now evaluate the the magnon dispersion and the inelastic structure
factor for even and odd excitations numerically. From a comparison of
our findings for the monolayer system to neutron scattering data for
the cuprate compounds we found\cite{Kru+03} the coupling $\lambda J$
across a stripe to be about one order of magnitude smaller than the
nearest neighbor coupling $J$ within the domains.  For the coupling
$\mu J$ between the layers a value $\mu\approx0.08$ is
reported\cite{Reznik+96} for antiferromagnetic YBCO in the absence of
stripes.  Therefore in the stripe system the couplings $\lambda$ and
$\mu$ can be assumed to be of the same order. In the following we keep
the value of $\lambda$ fixed and discuss the effects of increasing
$\mu$ starting from the case of decoupled layers ($\mu=0$) where the
band structure of the monolayer system\cite{Kru+03} should be
recovered. In this parameter regime the classical ground states for
shifted parallel and perpendicular charge stripes show the canted
planar texture and the antiphase domain boundary character of the
charge stripe is weakened by the interlayer coupling but still
pronounced.  Finally we shortly present the excitation spectra for
shifted parallel and perpendicular stripes for parameters belonging to
the collinear ground state regime.

\begin{figure}[h]
  \epsfig{figure=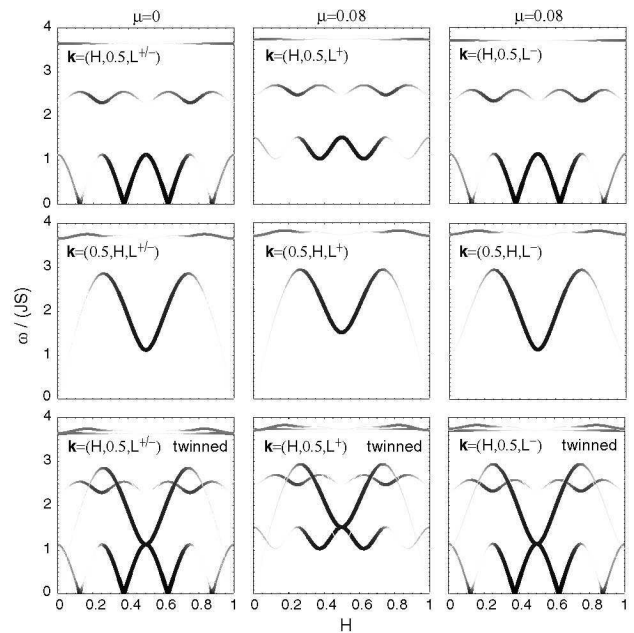,width=0.95\linewidth}
\caption{%
  Band structure and spectral weight along the $(H,0.5,L^\pm)$ and
  $(0.5,H,L^\pm)$ directions for parallel stripes lying on top of each
  other and couplings $\lambda=0.15$ and $\mu=0$, $0.08$. The last row
  shows the band structure of a twinned sample (see text).  $L^+$
  corresponds to even, $L^-$ to odd excitations.  Darker and larger
  points correspond to a larger weight of the inelastic structure
  factor.}
    \label{par1}
\end{figure}

\begin{figure}[h]
\epsfig{figure=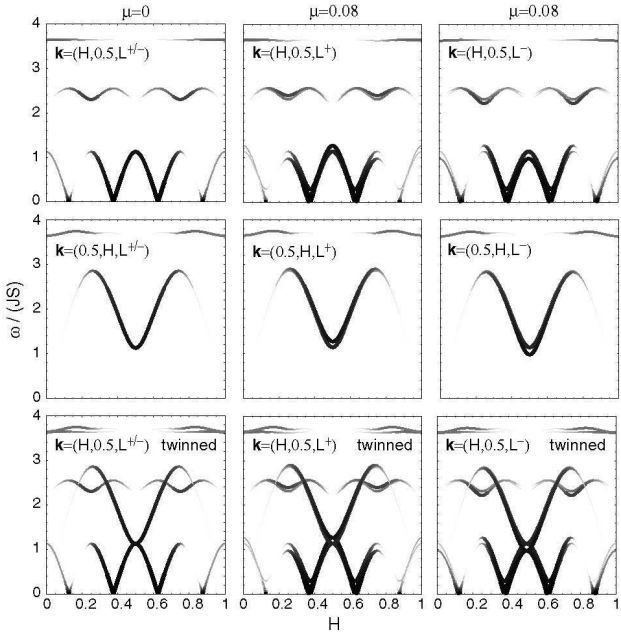,width=0.95\linewidth}
\caption{
  Band structure and spectral weight along the $(H,0.5,L^\pm)$ and
  $(0.5,H,L^\pm)$ directions for shifted parallel stripes and
  couplings $\lambda=0.15$ and $\mu=0,0.08$. The last row shows the
  resulting band structure of a twinned sample.}
  \label{par2}
\end{figure}

In the case of decoupled layers ($\mu=0$) the results of the
monolayer system are trivially recovered.  Since the two layers
are uncorrelated, the structure factor does not depend on the $L$
component of the wave vector.  For parallel stripes (with or
without a relative shift of the stripes) where the charge
modulation is unidirectional with $\qch_1=\qch_2=(1/4,0)$ we just
obtain an additional twofold degeneracy of each of the three bands
due to the equivalence of the two layers. Therefore the
degeneration of the bands is fourfold since in the monolayer case
each band is twofold degenerated due to the equivalence of the two
sublattices.\cite{Kru+03} The lowest, acoustical band has zeros at
the magnetic superstructure vectors which are located at $(j/4,0)$
and $(j/4+1/8,1/2)$, $j=0,\ldots,3$, within the Brillouin zone of
the square lattice (we choose $0\le H,K<1$). The spectral weight
is concentrated near the lowest harmonic incommensurate wave
vectors $\bQ=(1/2\pm 1/8,1/2)$. With increasing energy the
incommensurability decreases and the branches of the acoustic
magnon band close at the antiferromagnetic wave vector $(1/2,1/2)$
and an energy $\w_\pi$ which we associate with the
$\pi$-resonance. Along the $(H,1/2)$ direction the acoustic band
is gapped to the overlying optical magnon band (see upper left
panels in Figs.~\ref{par1} and \ref{par2}).  Along the orthogonal
direction $(1/2,K)$, one optical band has vanishing spectral
weight and only two bands are visible (see middle-left panels in
Figs.~\ref{par1} and \ref{par2}).

In twinned samples with stripe domains oriented orthogonal to each
other, a scan along the $(H,1/2)$ direction results in the
superposition of the signals obtained from scans in directions
$(H,1/2)$ and $(1/2,H)$ of a single-domain sample.  For domains of
equal size, one thus obtains an apparent symmetry $(H,K)
\leftrightarrow (K,H)$ and a fourfold pattern of the static
incommensurate wave vectors located at $\bQ=(1/2\pm 1/8,1/2)$ and
$\bQ=(1/2,1/2\pm 1/8)$ also for (shifted) parallel stripes.  In
Figs.~\ref{par1} and \ref{par2}, the panels in the third row are just
obtained by superimposing the panels of the first and second row.
Since the acoustic band of the monolayer system has a saddle point at
the antiferromagnetic wave vector, the resulting band structure is
x-shaped in the vicinity of the $\pi$-resonance energy.

The configuration of hole stripes lying perpendicular to each other
corresponds to charge modulation wave vectors $\qch_1=(1/4,0)$ and
$\qch_2=(0,1/4)$. For decoupled layers, the resulting band structure
contains the bands of the monolayer system and the same bands rotated
by 90 degrees leading to the symmetry $\w(H,K)=\w(K,H)$ and therefore
to a fourfold pattern of the static incommensurate wave vectors
located at $\bQ=(1/2\pm 1/8,1/2)$ and $\bQ=(1/2,1/2\pm 1/8)$.  Thus,
for $\mu=0$, the structure factor is identical for perpendicular
stripes and twinned parallel stripes (left lower panel in
Figs.~\ref{par1} and \ref{par2}).

\begin{figure}[h]
\epsfig{figure=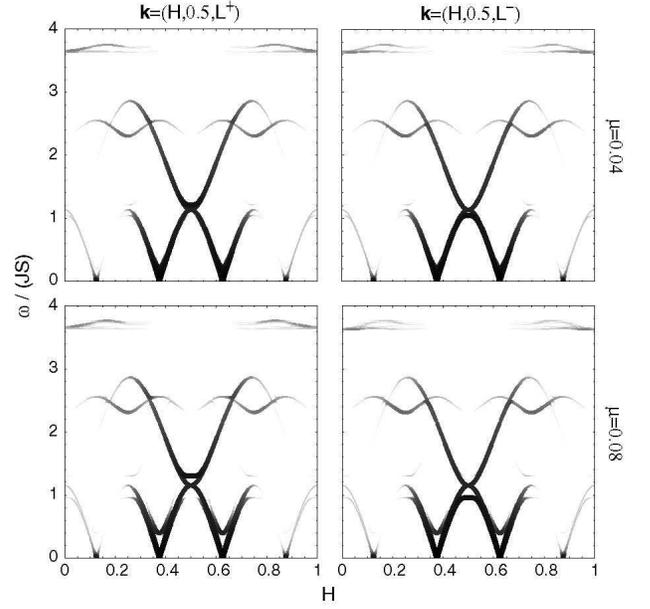,width=0.95\linewidth}
\caption{%
  Band structure and spectral weight for even (left panel) and odd
  (right panel) excitations along $(H,0.5,L^\pm)$ direction for
  perpendicular stripes with couplings $\lambda=0.15$ across the
  stripes and interlayer couplings $\mu=0.04$ (upper row) and
  $\mu=0.08$ (lower row).}
  \label{perp}
\end{figure}

With increasing interlayer coupling $\mu$ the bands start to split
with different distributions of the spectral weights in the odd and
even channel (cf. Figs.~\ref{par1}, \ref{par2}, and \ref{perp}).  For
parallel and shifted parallel stripes the Hamiltonian is invariant
under the reflection $z \to -z$ combined with a translation.  This
implies that the magnon states -- modulo a phase factor which does not
enter the structure factor -- have a well defined parity with respect
to an exchange of both layers.  As a consequence, nondegenerate bands
are visible only either in the even or the odd channel.

Nevertheless the excitation spectra of the two parallel stripe
configurations deviate significantly, e.g. the even excitations
are gapped for parallel stripes whereas for shifted parallel
stripes the intensity of even excitations is only reduced at low
energies (cf. middle columns of Figs.~\ref{par1}, \ref{par2}). For
stripes on top of each other each band -- which is fourfold
degenerate at $\mu=0$ -- splits up into twofold degenerate bands
which have identical parity. For shifted stripes each band splits
up into three bands. One of them is twofold degenerate and both
subbands are of opposite parity. Therefore this degenerated band
is visible in both channels (cf. Fig.~\ref{par2}).

For perpendicular stripes the symmetry is more complicated.  The
Hamiltonian is invariant under a reflection $z\to-z$ in
combination with a 90$^\circ$-rotation along the $z$ axis.  Since
this rotation mixes different wave vectors, almost all eigenstates
do not have a well defined parity and will be partially visible in
the odd and even channel.  The exception are modes at particular
wave vectors such as the antiferromagnetic wave vector which are
mapped onto themselves (modulo a reciprocal lattice vector). Only
there the excitations can be classified due to their symmetry.
Like for the shifted parallel stripes the excitations are not
gapped in the even channel (cf. Fig.~\ref{perp}).

\begin{figure}[h]
\epsfig{figure=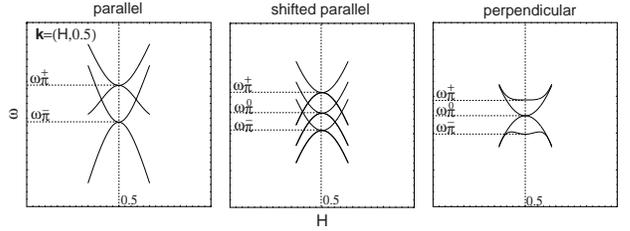,width=0.95\linewidth}
\caption{%
  Schematic illustration of the band splitting in the vicinity of the
  antiferromagnetic wave vector $(1/2,1/2)$ along the $(H,1/2)$
  direction. In the cases of parallel stripes the band structures for
  twinned samples are shown. Even and odd bands are gathered
  together.}
\label{splits}
\end{figure}

We now focus on the band splitting and the distribution of the
spectral weights of even and odd excitations at the antiferromagnetic
wave-vector $(1/2,1/2)$. With increasing interlayer coupling $\mu$,
the resonance energy $\w_\pi$ splits up into two different energies
$\w_\pi^-$ and $\w_\pi^+$ for centered-parallel stripes and into three
energies $\w_\pi^-$, $\w_\pi^0$ and $\w_\pi^+$ for the other stripe
configurations as schematically illustrated in Fig.~\ref{splits}.  It
is common to all stripe configurations that $\w_\pi^-$ has a finite
spectral weight only in the odd channel, whereas $\w_\pi^+$ has a
finite weight only in the even channel.  For shifted parallel and
perpendicular stripes, in both channels a finite intensity is found
at the intermediate energy $\w_\pi^0$.  This intensity is however
smaller than at $\w_\pi^\pm$.

\begin{figure}[h]
\epsfig{figure=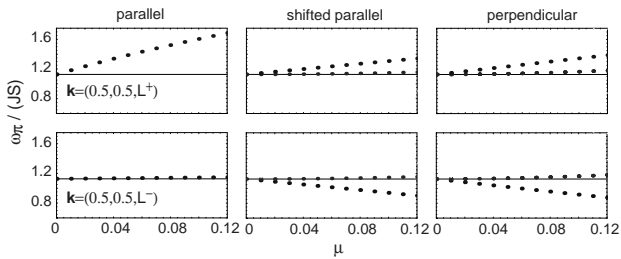,width=0.95\linewidth}
\caption{%
  Splitting of the resonance energy as a function of the interlayer
  coupling $\mu$ for $\lambda=0.15$. In the odd channel ($L=L^-$) the
  spectral weight is concentrated at $\w_\pi^-$ and no intensity is
  found at $\w_\pi^+$, in the even channel ($L=L^+$) no excitations at
  $\w_\pi^-$ are observable and the spectral weight is concentrated at
  $\w_\pi^+$. For shifted and perpendicular stripes in both channels a
  small intensity is found at the intermediate energy $\w_\pi^0$.}
  \label{split}
\end{figure}

The splitting of the resonance energy for shifted parallel and
perpendicular stripes looks quite similar.  $\w_\pi^-$ and $\w_\pi^+$
are almost equidistant to the intermediate energy $\w_\pi^0$ which
increases only slightly with $\mu$ (cf.  Fig.~\ref{split}). For small
couplings the splitting is quadratic in $\mu$. For centered-parallel
stripes the splitting looks different, $\w_\pi^+$ increases almost
linearly with $\mu$ whereas $\w_\pi^-$ is almost independent of the
interlayer coupling.

\begin{figure}[h]
\epsfig{figure=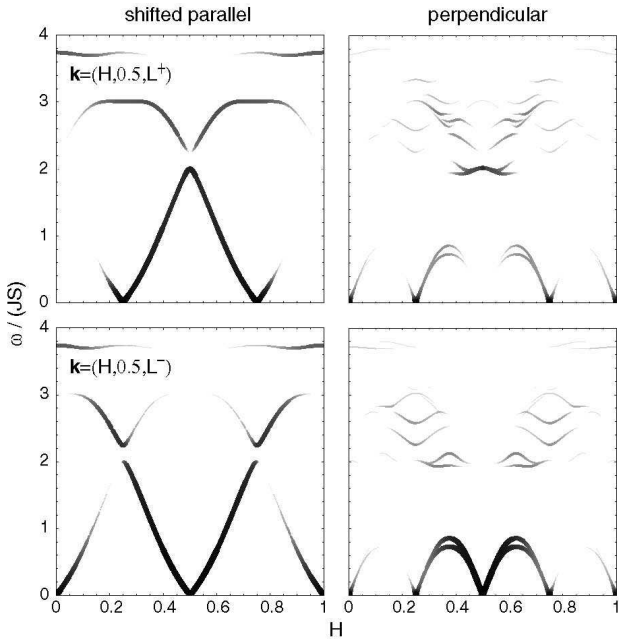,width=0.95\linewidth}
\caption{%
  Band structure in the collinear regime $\mu>\mu_c(\lambda)$ along
  the direction $(H,1/2,L^\pm)$ for shifted parallel and perpendicular
  stripes with $\lambda=0.07$ and $\mu=0.50$. }
  \label{coll}
\end{figure}

Finally, we calculate the band structures for shifted parallel and
perpendicular stripes for couplings $\lambda<\lambda_c$ and
$\mu>\mu_c(\lambda)$ where the ground-states are collinear and the
charge stripes lose their anti-phase domain boundary character. We
implicitly assume that $\mu$ is not too large, otherwise spins on
top of each other dimerize and lose their magnetization. In this
regime the magnetic fluctuations are drastically changed. For both
stripe orientations, the odd channel now has a static signal at
the antiferromagnetic wave vector, whereas in the even channel the
spectral weight is concentrated at incommensurate positions
$(1/2\pm 1/4,1/2)$ (cf. Fig.~\ref{coll}). For perpendicular
stripes we also find small intensity at this positions in the odd
channel. The incommensurability is doubled compared to the regime
of canted planar ground-states reflecting that the charge stripes
do not act like anti-phase domain boundaries in the regime of
strongly coupled layers.  In the even channel the intensity at the
antiferromagnetic wave-vector is peaked at an energy $\w_\pi$
which increases with the interlayer coupling $\mu$ and is
approximately the same for both stripe configurations.

\section{Discussion}
\label{sec.discussion}

In this section we compare our results to neutron scattering data for
the bilayer high-$T_c$ compound $\mathrm{YBa_2Cu_3O_{6+x}}$.  We wish
to stress that -- because of the simplifications assumed in our model
-- it is not our goal to obtain a quantitative agreement.  Rather we
wish to draw a qualitative comparison in order to fortify the
hypothesis that the stripe picture is a suitable approach to describe
spin fluctuations.  Furthermore, we hope that a comparison of future
experimental data with our calculations will help to identify the
realized stripe configuration.

Since a spin gap with an energy $\w_{\textrm{gap}}$ -- e.g. due to
Cooper-pair formation -- is not incorporated in our model, the results
apply only to energies above $\w_{\textrm{gap}}$ where the magnon
dispersion is not masked by the superconducting condensate. In
particular in the underdoped regime where $\w_{\textrm{gap}}$
decreases with the doping level, the calculated spectral features
become visible over an increasing energy range.  Our calculations are
restricted to zero temperature.  Therefore, a comparison can also be
made only to experiments performed at temperatures well below the
superconducting transition temperature.

Experiments\cite{Mook+00,Stock+02,Stock+03} in (partially) detwinned
YBCO provide evidence for unidirectional order, i.e., that a fourfold
pattern of incommensurate peaks near the antiferromagnetic wave vector
$\bkaf=(1/2,1/2)$ results only from the twinning.  The stripes seem to
be parallel and oriented along the direction of the oxygen chains in
the adjacent planes.  This immediately speaks against the scenario of
perpendicular stripes for which detwinning would not affect the
fourfold symmetry.

We briefly recall some neutron scattering measurements on
$\mathrm{YBa_2Cu_3O_{6+x}}$ which provide insight into the
incommensurability and the $\pi$ resonance over a wide doping and
temperature range. It was controversial for quite some time whether
both phenomena would exist above $T_c$ until in underdoped materials
the incommensurability was found also above $T_c$.\cite{Dai+98}
Likewise, the appearance of the magnetic resonance was found above
$T_c$, occurring together with the pseudogap at a temperature
$T^*>T_c$ determined from transport and nuclear
resonance.\cite{Dai+99} Although the $\pi$-resonance persists as a
well defined feature also in the normal state above $T_c$, its
intensity can be reduced significantly at $T_c$.\cite{Stock+03} For
near optimally doped compounds, the resonance is not detectable in the
normal phase\cite{Bourges+00} since $T^*$ almost coincides with $T_c$.
Dai \emph{et al.}\cite{Dai+01} concluded that the resonance exists
above $T_c$ for $x\le 0.8$ and that incommensurate spin fluctuations
appear in the normal state for $x\le0.6$.  Arai \emph{et
  al.}\cite{Arai+99} also observed incommensurate fluctuations in the
normal state for a sample with an oxygen concentration of $x=0.7$.
Thus, superconductivity is not a prerequisite for incommensurability
and $\pi$ resonance in bilayer compounds as well as in monolayer
compounds.

For underdoped YBCO with various oxygen concentrations, the
experimentally observed spin dynamics data (see Table~\ref{mat}) look
qualitatively very similar.  There is a systematic increase of the
incommensurability and of the $\pi$ resonance frequency with doping,
which is consistent with our model. We have shown this recently for a
monolayer model.\cite{Kru+03} The bilayer stripe model shares this
feature and therefore we focus in this paper exclusively on specific
bilayer features.

\begin{table}[htbp]
  \begin{center}
    \begin{tabular}{c||c|c|c|c|c|c|c|c}
      $x$ & $0.35$ & $0.45$ & $0.5$ & $0.5$ & $0.6$ & $0.7$ &
      $0.7$& $0.7$
      \\\hline
      $T_c$ (K)& $39$ & $48$ & $52$ & $59$ & $63$ & $67$ &
      $67$& $74$
      \\\hline
      $\delta$ (r.l.u.)& $1/16$ &  & & $0.08$
       & $0.10$ & $1/8$ & & $0.1$
       \\\hline
      $p$ & 8 & & & $6.25$ & $5$ & $4$ & & $5$
      \\\hline
      $\w_\pi^-$ (meV) & $23$ & $30.5$ & $31.5$ & $33$ & $34$ & $36$ &
      $33$&$37$
      \\\hline
      $\w_\pi^+$ (meV) & & & & & & $41$ & $50$ &
      \\\hline
      Ref. & \onlinecite{Mook+02} & \onlinecite{Dai+01} &
      \onlinecite{Dai+01} & \onlinecite{Stock+03} &
      \onlinecite{Dai+01}  & \onlinecite{Arai+99}& \onlinecite{Fong+00} & \onlinecite{Dai+01}
    \end{tabular}
  \end{center}
\caption{%
  Spin dynamics data for $\mathrm{YBa_2Cu_3O_{6+x}}$ for
various  oxygen concentrations $x$ characterized by the critical
temperature $T_c$, incommensurability $\delta$, corresponding
stripe period $p$, the resonance energy $\w_\pi^-$ observed in the
odd channel, and $\w_\pi^+$.} \label{mat}
\end{table}

Experimentally, constant energy scans slightly above the gap in
the odd channel along $(H,1/2,L^-)$ show a broad intensity peak at
$\bkaf$, before incommensurate scattering sets in and the data can
be compared to our model. The intensity shows magnetic peaks at a
distance $\dk(\w)$ away from $\bkaf$.  The incommensurability
$\delta$ is determined by extrapolating $\dk(\w)$ to $\w=0$ and it
is connected to the stripe spacing $p$ through $\delta=1/(2p)$.
The incommensurate peaks are best defined if the stripe spacing is
nearly a multiple of the lattice spacing (integer $p$) since the
stripes are stabilized by the lattice.\cite{Mook+02}

The three stripe configurations examined for our model are not
equivalent in their low-energy behavior. For (unshifted) parallel
stripes (see Fig.~\ref{par1}), an incommensurability is visible at low
energies only in the odd channel since the even channel has a
relatively large gap not related to superconductivity.  In contrast,
for shifted parallel and perpendicular stripes the even channel shows
incommensurate response down to the superconducting gap.  Experimental
evidence\cite{Bourges+97prb,Dai+99,Fong+00} for a large gap in the
even channel (well above the resonance energy in odd channel)
therefore favors the configuration with unshifted parallel stripes.

With increasing energy, the separation $\dk(\w)$ of the incommensurate
peaks decreases and the branches close at $\bkaf$ at certain energies
$\w_\pi$. Depending on the stripe configuration, there are two or
three such energies, compare Fig.~\ref{splits}.  According to our
model, an energy scan of the odd channel at $\bkaf$ would show a first
resonance at the intersection with the lowest magnon band at
$\w_\pi^-$ which we identify with the resonance
frequency.\cite{Rossat-Mignod+91,Rossat-Mignod+93} For shifted
parallel and perpendicular stripes, a second line at $\w_\pi^0$
contributes to the odd channel.  It has significantly less weight and
is separated from the first one by only a small energy splitting (of
the order of a few meV) which would be hard to be resolved
experimentally.

In a similar way, the even channel has a resonance at an energy
$\w_\pi^+>\w_\pi^-$, and for shifted parallel an perpendicular stripes
also a weaker resonance at an intermediate frequency $\w_\pi^0$ (cf.
Fig.~\ref{splits}).
Experimentally,\cite{Rossat-Mignod+93,Mook+93,Stock+03} a strong
oscillatory dependence of the scattering intensity on $L$ shows that
the resonance frequencies in the odd and even channel are well
separated. Energy scans at the antiferromagnetic wave vector show
peaks at $\w_\pi^-$ in the odd channel and $\w_\pi^+$ in the even
channel, no peak at the intermediate energy $\w_\pi^0$ which should be
visible in both channels is resolved.\cite{Fong+00} This again favors
unshifted parallel stripes, which (in contrast to shifted parallel and
perpendicular stripes) have no shared resonance frequency $\w_\pi^0$.
Although we restricted our comparison to experiments on underdoped
samples, overdoped compounds also show two distinct resonance modes of
opposite symmetry,\cite{Pailhes+03} which could be identified with
$\w_\pi^-$ and $\w_\pi^+$.

From a comparison of the band splitting
$\Delta\w_\pi=\w_\pi^+-\w_\pi^-$ to experimental values (cf.
Tab.~\ref{mat}) we can estimate the strength of the interlayer
coupling $\mu$.  For $\lambda=0.15$ we find $\mu\approx0.02-0.06$
almost independent of the stripe configuration.  This value is
reasonable since the effective coupling $\mu$ in the stripe system
should be slightly reduced compared to the undoped case where a value
of $\mu\approx0.08$ is reported.\cite{Reznik+96}

Above $\w_\pi$ the response is found to become incommensurate again
with increasing separation $\dk(\w)$.  The momentum width is larger
and the intensity is weaker than below $\w_\pi$.  Overall, the
dispersion is ``x-shaped''.  As pointed out in Sec.~\ref{sec.results}
such a shape appears basically for every non-unidirectional stripe
configuration, for parallel stripes in twinned crystals as well as in
perpendicular stripes.  The x-shape has been observed explicitly in
Refs.~\onlinecite{Bourges+97prb,Arai+99,Fong+00,Mook+02}.  It would be
interesting to verify in detwinned samples that the relative
intensities of the upper and lower branches of the x-shape are related
to the population ratio of the twin domains.

In conclusion, we have calculated the bilayer effects in the magnetic
excitation spectrum in striped states.  As a generic feature of the
stripe model we find x-shaped dispersion relation in the vicinity of
the $\pi$ resonance, which is consistent with experimental data.  We
have obtained a bilayer splitting of single-layer bands into two or
three bilayer bands.  From the three stripe configurations studied,
the unshifted parallel case overall is most consistent with neutron
scattering data, although it seems to be energetically unfavorable at
first sight.

\begin{acknowledgments}
  We gratefully acknowledge helpful discussions with M. Braden, Y.
  Sidis, and J.M.  Tranquada.  This project was supported financially
  by Deutsche Forschungsgemeinschaft (SFB608).
\end{acknowledgments}

%\bibliographystyle{apsrev}
%\bibliography{strings,refs}

\end{document}